\documentclass[fleqn,usenatbib]{mnras}

\usepackage{newtxtext,newtxmath}

\usepackage[T1]{fontenc}

\DeclareRobustCommand{\VAN}[3]{#2}
\let\VANthebibliography\thebibliography
\def\thebibliography{\DeclareRobustCommand{\VAN}[3]{##3}\VANthebibliography}

\usepackage{graphicx}	
\usepackage{amsmath}	
\usepackage[normalem]{ulem}  

\newcommand{\WI}[2]{#1_{\mathrm{#2}}}
\newcommand{\isn}[2]{\mbox{$^{#2}${#1}}}
\newcommand{\Ye}{\WI{Y}{e}}
\newcommand{\ad}[1]{#1}
\newcommand{\add}[1]{#1}

\title[Neutrinos from pre-supernova]{Neutrinos from pre-supernova in the framework of TQRPA method}

\author[A. A. Dzhioev, A. V. Yudin, N. V. Dunina-Barkovskaya, A. I. Vdovin]{
A. A. Dzhioev$^{1}$\thanks{E-mail: dzhioev@theor.jinr.ru}, A. V. Yudin$^{2,3}$\thanks{E-mail: yudin@itep.ru}, N. V. Dunina-Barkovskaya$^{2}$, A. I. Vdovin$^{1}$
\\
$^{1}$ Bogoliubov Laboratory of Theoretical Physics, Joint Institute for Nuclear Research, 141980 Dubna, Russia\\
$^{2}$ National Research Center Kurchatov Institute, pl. Kurchatova 1, Moscow, 123182, Russia\\
\add{$^{3}$ Novosibirsk State University, Pirogova, 2, 630090 Novosibirsk, Russia}
}

\date{Accepted XXX. Received YYY; in original form ZZZ}

\pubyear{2023}

\begin{document}
\label{firstpage}
\pagerange{\pageref{firstpage}--\pageref{lastpage}}
\maketitle

\begin{abstract}
We propose a new method for calculating spectra and luminosities for (anti)neutrinos produced in the pre-supernova environment by weak processes with hot nuclei. It is based on the thermal quasiparticle random phase approximation (TQRPA), that allows microscopic thermodynamically consistent
calculations of the weak-interaction  response of nuclei at finite temperatures. For realistic representative pre-supernova conditions from the stellar evolution code MESA, we compute (anti)neutrino luminosities and spectra arising from neutral- and charged-current weak reactions with hot $^{56}$Fe and compare them with the contribution of thermal processes. We find that the TQRPA approach produces not only a
higher total luminosity of electron neutrinos (mainly born in the
electron capture reaction), compared to the standard technique   based on the large-scale shell model \ad{(LSSM)} weak-interaction rates, but also a  harder neutrino spectrum.  Besides, \ad{applying the TQRPA and LSSM,}  we find that in the context of electron antineutrino generation, the neutral-current nuclear de-excitation (ND) process via neutrino-antineutrino pair emission is at least as important  as the electron-positron pair annihilation process. We also show that flavor oscillations enhance the high-energy contribution of the ND process to the \ad{electron} antineutrino flux. This could potentially be important for pre-supernova antineutrino registration by the Earth's detectors.
\end{abstract}




\section{Introduction}

Supernova explosions are magnificent and not yet fully understood phenomena of nature. For dozens of days the dying star shines like an entire host galaxy. But despite this, in massive stars that end their lives as a core-collapse supernova, the main amount of energy during the explosion (of an order of a few $10^{53}$ erg) is lost by neutrinos. The first and so far the only detection of neutrinos from a supernova occurred on February 23, 1987; it was a famous supernova SN1987A that flared up in the Large Magellanic Cloud (see, e.g., \cite{Hirata1987PhRvL,Bionta1987PhRvL, Alekseev1987JETPL, Ryazhskaya2006PhyU}).

But even before the explosion, neutrino plays an important role in the life of a massive star: already at the stage of the neon-oxygen core (a few years before the collapse), the neutrino luminosity becomes higher than the photon luminosity \citep{Weaver1978ApJ}, and just before the collapse  the energy loss via neutrino emission is many orders of magnitude superior to other processes~\citep{Kato2017ApJ}. Such a star, called a pre-supernova, appears to be a promising source of neutrinos to be detected \citep{Odrzhywolek2004AcPPB,Patton2017ApJ_2}. Pre-supernova (anti)neutrinos, if registered, can serve as an alarm signal of an upcoming explosion \citep{Kutschera2009AcPPB}. Moreover, since neutrinos can freely propagate through the stellar matter, they carry direct information about  thermodynamic conditions in the core and their observation would offer a possibility for studying the physical processes that lead to the core collapse.

Neutrino emission in a stellar core occurs due to a number of thermal processes and nuclear weak reactions.  Thermal neutrinos are mostly emitted via the electron-positron pair annihilation and plasmon decay, and  their production  is entirely determined by the  core temperature, density and electron fraction  \citep{Itoh1996, Kato2015ApJ}. In contrast, neutrino emission via nuclear weak reactions, such as $\beta^\pm$-decay and $e^\pm$-capture, has a stronger dependence on the isotopic composition of the core and its study therefore  requires realistic stellar evolution simulations with a large and accurate nuclear reaction network \citep{Odrzywolek2010AcPPB}.  However, not only isotopic composition but also thermodynamic conditions in the core  affect the  neutrino production in nuclear weak reactions.
The reason is that    nuclear excited states are thermally populated in the  high-temperature stellar environment and weak reactions on thermally excited states allow the emission of neutrinos with rather high energy. The relative contribution of these high-energy neutrinos to the total flux depends on temperature. It is important to realize that despite the Boltzmann suppression, the gain in phase space combined with  the large matrix element and increased density of \ad{thermally} excited states   can make this contribution significant.
Eventually, the balance between all these factors  depends on thermodynamic conditions and properties of individual nuclei.

In \cite{Patton2017ApJ_1,Patton2017ApJ_2}, the contribution of charged-current nuclear weak processes to neutrino emissivities was analyzed for different  realistic thermodynamic conditions and isotopic compositions of the star obtained from the state-of-the-art stellar evolution code MESA. It was found that, under certain conditions, nuclear processes compete with thermal processes  in their contribution to the (anti)neutrino flux or even dominate in the energy window relevant for detection, which is mostly due to the decay and electron capture on $pf$-shell nuclei with $A=50-60$ (iron-group nuclei). However, it was pointed out that while total emissivities are relatively robust, the high-energy tails of the neutrino
spectra, in the detectable window, are very sensitive to the details of the calculations. Specifically, the source of the error lies in the so called single-strength (or effective $q$-value) approximation \citep{Langanke_PRC64}, which was adopted in \cite{Patton2017ApJ_1,Patton2017ApJ_2} to generate charged-current neutrino spectra.
An exploratory study of this error was performed in~\cite{Misch_PRC94}  and it was shown that the specific neutrino energy spectrum obtained from the single-strength approximation could miss important features when the most important weak transitions involve \ad{thermally} excited states.

In~\cite{Misch_PRC94}, (anti)neutrino emission  via charged-current  reactions  was considered using some  $sd$-shell nuclei as an example. To do this, shell-model calculations were preformed to obtain strength  distributions of Gamow-Teller and Fermi transitions which dominate nuclear weak reactions at  pre-supernova conditions. However, for  $pf$-shell nuclei   large-scale shell-model (LSSM) calculations  of the GT strength distributions require a huge dimension of the model space and, therefore,  they  are still limited by the nuclear ground and low-lying excited states~\citep{Caurier_RevModPhys77}.  When performing shell-model calculations of stellar weak-interaction rates,  this issue is overcome by employing the Brink hypothesis  and the method of "back-resonances"~\citep{Langanke2000NuPhA}.  However,  even if we could perform precise LSSM calculations for highly excited states, the state-by-state consideration of individual  contributions would remain  computationally infeasible  because of too many thermally populated states at pre-supernova temperatures  $T\approx 1$\,MeV \ad{($10^{10}\,\text{K}=0.86$\,MeV)}.

To avoid these shortcomings and compute stellar weak-interaction rates for hot \ad{thermally} excited nuclei in a microscopic thermodynamically consistent way, the thermal quasiparticle random-phase approximation (TQRPA) was proposed in~\cite{Dzhioev_PRC81}. The TQRPA approach  is based on  a statistical description of the hot nucleus and it enables one to obtain temperature dependent strength functions for nuclear transitions involved in the considered weak reaction. The advantage of the TQRPA is that it makes it possible to treat both endoergic and exoergic weak processes without application of the Brink hypothesis and the "back-resonance" method and without violation of the detailed balance principle. Later on, the combination of the TQRPA and the Skyrme energy density functional methods was successfully applied to study both charged-current~\citep{Dzhioev_PRC100_2, Dzhioev_PRC101} and  neutral-current~\citep{Dzhioev_PRC94, Kondratyev_PRC100_4} stellar weak reactions with $pf$-shell and heavier neutron rich nuclei (see also recent reviews \cite{Dzhioev_PhPN53_1,Dzhioev_PhPN53_2,Dzhioev_PhPN53_3}).

In our recent paper~\cite{Dzhioev_Particles6}, the TQRPA method was applied for the first time for computing  (anti)neutrino spectrum and energy loss rates due to nuclear weak processes under specific pre-supernova conditions which correspond to a realistic pre-supernova model. Using the hot \isn{Fe}{56} as an example, we have shown that thermodynamically consistent TQRPA calculations predict an electron neutrino spectrum  with an enhanced high-energy fraction. These high-energy neutrinos are emitted via the exoergic electron capture, i.e. the process when a thermally excited hot nucleus captures an electron and then de-excites by energy  transfer  to the emitted neutrino.  Another interesting process considered in~\cite{Dzhioev_Particles6} is the de-excitation of a hot nucleus via the neutrino-antineutrino pair emission.  \ad{This neutral-current process produces $\nu\bar\nu$-pairs of all flavors with equal probability and its potential importance}  for production of  high-energy pre-supernova (anti)neutrinos  was outlined in \citep{Misch_PRC94, Patton2017ApJ_2}.
\ad{In~\cite{Dzhioev_Particles6},   performing TQRPA calculations for both charged- and neutral-current weak-interaction  processes with $^{56}$Fe generating electron (anti)neutrinos,  we have shown that   $\nu\bar\nu$-pair emission has practically no effect on the $\nu_e$ spectrum, while this process  produces  more high-energy electron antineutrinos    than other nuclear weak processes (positron capture and $\beta^{-}$-decay).}

\ad{We  would also like to  mention  that to date the only study analyzing the impact of $\nu\bar\nu$-pair emission from nuclear de-excitation on  core-collapse supernova simulations has been performed by~\cite{Fischer_PRC88}. In particular, this study showed that   $\nu\bar\nu$-pair emission   has basically no impact on the global supernova properties and the energy loss during collapse is dominated by electron neutrinos produced by electron captures on nuclei.  However, it was found that nuclear de-excitation is the leading source for the production of electron antineutrinos as well as ($\mu,\,\tau$)  (anti)neutrinos during the collapse phase. }

In~\cite{Dzhioev_Particles6}, (anti)neutrino spectra and energy loss rates due to nuclear processes were computed at sample points inside the stellar core just before the onset of the collapse. At this moment, for the considered pre-supernova model, the iron isotope \isn{Fe}{56} dominates the isotopic composition of the central part of the stellar core.  The present work extends our previous study, and here we apply the TQRPA method at the same pre-supernova conditions  to calculate total (anti)neutrino spectra and  luminosities  generated in different nuclear weak processes and integrated over the volume of emission. In this paper, we also compute (anti)neutrino spectra and luminosities from  thermal processes  and take into account oscillation effects.

Our paper is organized as follows: In the next section, we briefly  outline the TQRPA method for computing neutrino spectra produced by hot nuclei in stellar interior. A comprehensive description of the TQRPA  and its application to study stellar weak-interaction processes is given in the recent reviews~\citep{Dzhioev_PhPN53_1,Dzhioev_PhPN53_2,Dzhioev_PhPN53_3}.   Then in Section~\ref{sec_PreSN}, we describe the structure of a particular pre-supernova model which has been used  to explore the role of weak nuclear processes in (anti)neutrino production. In  Section~\ref{sec_generation}, \ad{taking into account both nuclear and thermal processes,} we calculate  luminosities of different (anti)neutrino flavors inside the star and corresponding  spectra. After that\ad{, in Section~\ref{sec_features},} we compare the properties of \ad{(anti)neutrino spectra} obtained in the framework of the TQRPA with those obtained using a more conventional method based on the LSSM calculations \ad{and the effective $q$-value approximation}. Before proceeding to a discussion of the detectability of the calculated antineutrino fluxes
(see Section~\ref{sec_Detection} ),  in Section~\ref{sec_oscill}, we consider the issue of neutrino oscillation effects. Finally, we provide a short "Summary and Perspectives"\,  section.


\section{TQRPA method}

In what follows, we assume that under the pre-supernova conditions we are concerned, weak processes with  $pf$-shell nuclei are dominated by Gamow-Teller transitions and the emitted (anti)neutrinos freely leave the star. Then, the (anti)neutrino  spectra  for a single hot nucleus can be expressed in terms of  the temperature-dependent GT strength functions:
\begin{itemize}
    \item electron (EC) or positron (PC) capture
    \begin{align}\label{capture}
    &\phi^\text{EC,PC}(E_\nu)=\frac{G^2_\text{F} V^2_{\text{ud}} (g^*_A)^2}{2\pi^3\hbar^7c^6}\, E^2_\nu\times
    \notag\\
    &\int\limits^{+\infty}_{-E_\nu+m_ec^2}S_{\text{GT}_\pm}(E,T)E_ep_ec f_{e^\mp}(E_e)F(\pm Z, E_e)dE,
\end{align}
where the upper (lower) sign corresponds to EC (PC) and $E_e=E_\nu+E$, $p_ec = (E_e^2-m^2_ec^4)^{1/2}$;
\item $\beta^\mp$-decay
\begin{align}\label{decay}
    &\phi^{\beta^\mp}(E_\nu)=\frac{G^2_\text{F} V^2_{\text{ud}} (g^*_A)^2}{2\pi^3\hbar^7c^6}\, E^2_\nu\times
    \notag\\
    &\int\limits_{-\infty}^{-E_\nu-m_ec^2}S_{\text{GT}_\mp}(E,T)E_ep_ec (1-f_{e^\mp}(E_e))F(\pm Z+1, E_e)dE,
\end{align}
where the upper (lower) sign corresponds to the $\beta^-$- ($\beta^+$-) decay and $E_e=-E_\nu-E$;
\item $\nu\bar\nu$-pair emission via nuclear de-excitation (ND)
\begin{align}\label{pair}
    \phi^\text{ND}(E_\nu)=\frac{G^2_\text{F} g_A^2}{2\pi^3\hbar^7c^6}\, E^2_\nu
    \int\limits_{-\infty}^{\-E_\nu} S_{\text{GT}_0}(E,T)(E+E_\nu)^2dE.
\end{align}
Note that the ND process produces the same spectra for $\nu_e$ and $\bar\nu_e$. Moreover, the spectrum of other (anti)neutrino flavors is also given by~\eqref{pair}.
\end{itemize}

In the above expressions,  $G_\text{F}$ denotes the Fermi coupling constant,  $V_{\text{ud}}$ is the up-down element of the Cabibbo-Kobayashi-Maskava quark-mixing matrix and $g_A=-1.27$ is the weak axial coupling constant. Note that for charged-current reactions we use the effective coupling constant $g^*_A$ that takes into account the experimentally observed quenching of the GT$_\pm$ strength. The function  $f_{e^-(e^+)}(E)$ is  the Fermi-Dirac distribution for electrons (positrons) \ad{which depends on the chemical potential $\mu_{e^-}$ ($\mu_{e^+}=-\mu_{e^-}$) and temperature $T$,} and the Fermi function $F(Z,E)$ takes into account the distortion of the charged lepton wave function by the Coulomb field of the nucleus.

The  presented expressions for $\phi^i(E_\nu)$ ($i = \text{EC,\,PC,\,ND},\,\beta^\pm$) are exact in the sense that no additional assumptions or approximations are made in their derivation. Approximations should be made for computing the temperature dependent GT  strength functions defined as
\begin{equation}\label{str_funct1}
  S_{\mathrm{GT}_{\pm, 0}}(E,T) = \sum_{i,f} p_i(T)B^{(\pm,0)}_{if}\delta(E-E_{if}).
\end{equation}
Here $p_i(T)= e^{-E_i/kT}/Z(T)$ is the Boltzmann population factor for a state $i$ in the parent nucleus,  $B^{\pm,0}_{if} = |\langle f\|\mathrm{GT}_{\pm, 0}\|i\rangle|^2/(2J_i+1)$ is the reduced transition probability (transition strength) from the state $i$ to the state $f$ in the daughter nucleus; $\mathrm{GT}_0=\vec{\sigma}t_0$  for neutral-current reactions and $\mathrm{GT}_\mp=\vec{\sigma}t_\pm$ for
charged-current reactions\footnote{The zero component of the isospin operator is denoted by $t_0$, while $t_-$ and $t_+$ are the isospin-lowering ($t_-|n\rangle = |p\rangle$) and isospin-rising ($t_+|p\rangle = |n\rangle$) operators.}; $E_{if}$ is the transition energy between the initial and final states,  and it can be both positive and negative.
The pre-supernova model considered in this study has typical temperatures inside the star  around $T_9\approx1-10$ (see  the discussion below). Obviously,  at such high temperatures  an explicit state-by-state evaluation of the sums in Eq.~\eqref{str_funct1} is impossible with current nuclear models. We compute the  temperature-dependent strength function \eqref{str_funct1} applying the TQRPA approach which is a technique based on the quasiparticle random phase approximation extended to the
finite temperature by the superoperator formalism in the Liouville space~\citep{Dzhioev_PhPN53_1}.

 The central concept of the TQRPA is the thermal vacuum
$|0(T)\rangle$, a pure state in the  Liouville space, which corresponds to the grand canonical density matrix operator for a hot nucleus.
The time-translation operator in the Liouville space is the so-called thermal Hamiltonian $\mathcal{H}$ constructed from the nuclear Hamiltonian after introducing the particle creation and annihilation superoperators. Within the TQRPA,  the  strength function~\eqref{str_funct1} is
expressed in terms of the transition matrix elements from the thermal
vacuum to the eigenstates (thermal phonons) of the thermal Hamiltonian $\mathcal H|Q_i\rangle= \omega_i|Q_i\rangle$:
\begin{equation}\label{str_funct2}
 S_{\mathrm{GT}_{\pm, 0}}(E,T) = \sum_i \mathcal{B}^{(\pm,0)}_{i}\delta(E-\omega_i \mp \Delta_{np}).
\end{equation}
Here $\mathcal{B}^{(\pm,0)}_{i} = |\langle Q_i\|\mathrm{GT}_{\pm, 0}\|0(T)\rangle|^2$ is the transition strength to the $i$th thermal phonon state of a hot nucleus and $E_i^{(\pm, 0)} = \omega_i \pm \Delta_{np}$ is the respective transition energy; $\Delta_{np}=0$ for charge-neutral transitions, while  for charge-exchange transitions $\Delta_{np}=\delta\lambda_{np} + \delta M_{np}$, where $\delta\lambda_{np} = \lambda_n-\lambda_p$ is the difference between neutron and proton chemical potentials in the nucleus, and $\delta M_{np}=1.293$\,MeV is the neutron-proton mass splitting. Note that eigenvalues of the thermal Hamiltonian, $\omega_i$, can be both positive and negative. As a result, the strength functions~\eqref{str_funct2} \ad{for upward ($E>0$) and downward ($E<0$) transitions} obey the detailed balance principle:
\begin{equation}\label{DB1}
   S_{\mathrm{GT}_{0}}(-E,T)= \mathrm e^{-E/kT}  S_{\mathrm{GT}_{0}}(E,T)
\end{equation}
for charge-neutral GT transitions, and
  \begin{equation}\label{DB2}
   S_{\mathrm{GT}_{\mp}}(-E,T)= \mathrm e^{-(E\mp\Delta_{np})/kT}  S_{\mathrm{GT}_{\pm}}(E,T)
\end{equation}
for charge-exchange GT transitions. This property makes the TQRPA approach thermodynamically consistent. Due to  negative-energy \ad{downward GT transitions}, the $\beta^\pm$-decay and $\nu\bar\nu$-pair emission become possible for hot nuclei which are stable in their ground-state. Moreover, due to enhanced phase space such negative-energy transitions  may dominate the EC and PC processes.

In~\cite{Dzhioev_Particles6}, (anti)neutrino spectra due to weak processes with  hot \isn{Fe}{56} were computed applying  self-consistent TQRPA calculations based on the SkM* parametrization  of the Skyrme effective nucleon-nucleon interaction. Here "self-consistent"{} means that both the mean field for protons and neutrons and the residual interaction are obtained
from the same  energy density functional. In the present study, we  also use the  SkM* parametrization. To take into account the quenching of the GT$_\pm$ strength in $^{56}$Fe, we use the effective coupling constant $g^*_A = 0.56 g_A$.  Then, the total GT$_\pm$ strength in $^{56}$Fe at $T=0$  becomes close to the experimental one and to the LSSM results~\citep{Caurier_NPA653}.

Before proceeding further, let us briefly recall the main results obtained in~\cite{Dzhioev_Particles6}:

\begin{itemize}
    \item Considering temperature evolution of the GT strength functions in \isn{Fe}{56},  we have shown that the TQRPA does not support the Brink hypothesis and under relevant pre-supernova conditions thermal population of nuclear excited states makes possible (unblocks) negative- and low-energy GT transitions whose strength increases with temperature.  Negative-energy \ad{downward} transitions contribute to exoergic weak reactions with hot nuclei.

   \item Calculated \ad{electron} neutrino spectra confirm the conclusion of~\cite{Langanke_PRC64} that the  \ad{effective $q$-value} approximation can be applied under stellar conditions  with the electron chemical potential high enough to allow  the excitation of the GT$_+$ resonance by electron capture. Such conditions occur during the collapse phase. However, our calculations clearly demonstrate that this approximation can fail in the pre-supernova phase when negative-energy GT$_+$ transitions from thermally excited states noticeably contribute to electron capture and the resulting neutrino energy spectrum is double-peaked. On the whole, thermodynamically consistent calculations of the electron neutrino spectra  performed without assuming the Brink hypothesis  indicate that thermal effects on the GT$_+$ strength function  shift the spectrum to higher energies, and thus make \ad{$\nu_e$}
   neutrino detection more likely.

   \item The inclusion into consideration of the $\nu\bar\nu$-pair emission by hot nuclei  shows that this neutral-current process might be a dominant source of high-energy \ad{($E_{\bar\nu}\approx5-10$\,MeV) $\bar\nu_e$ and ($\mu,\,\tau$)} (anti)neutrinos emitted via de-excitation of the GT$_0$ resonance.  Moreover, since the ND process  is entirely independent of the electron  density and only depends on temperature, the detection of high-energy pre-supernova antineutrinos might be a test for thermodynamic conditions in the stellar interior.
\end{itemize}

\section{Pre-sn structure}\label{sec_PreSN}

\begin{figure*}
	\begin{center}
		\includegraphics[width=1\textwidth]{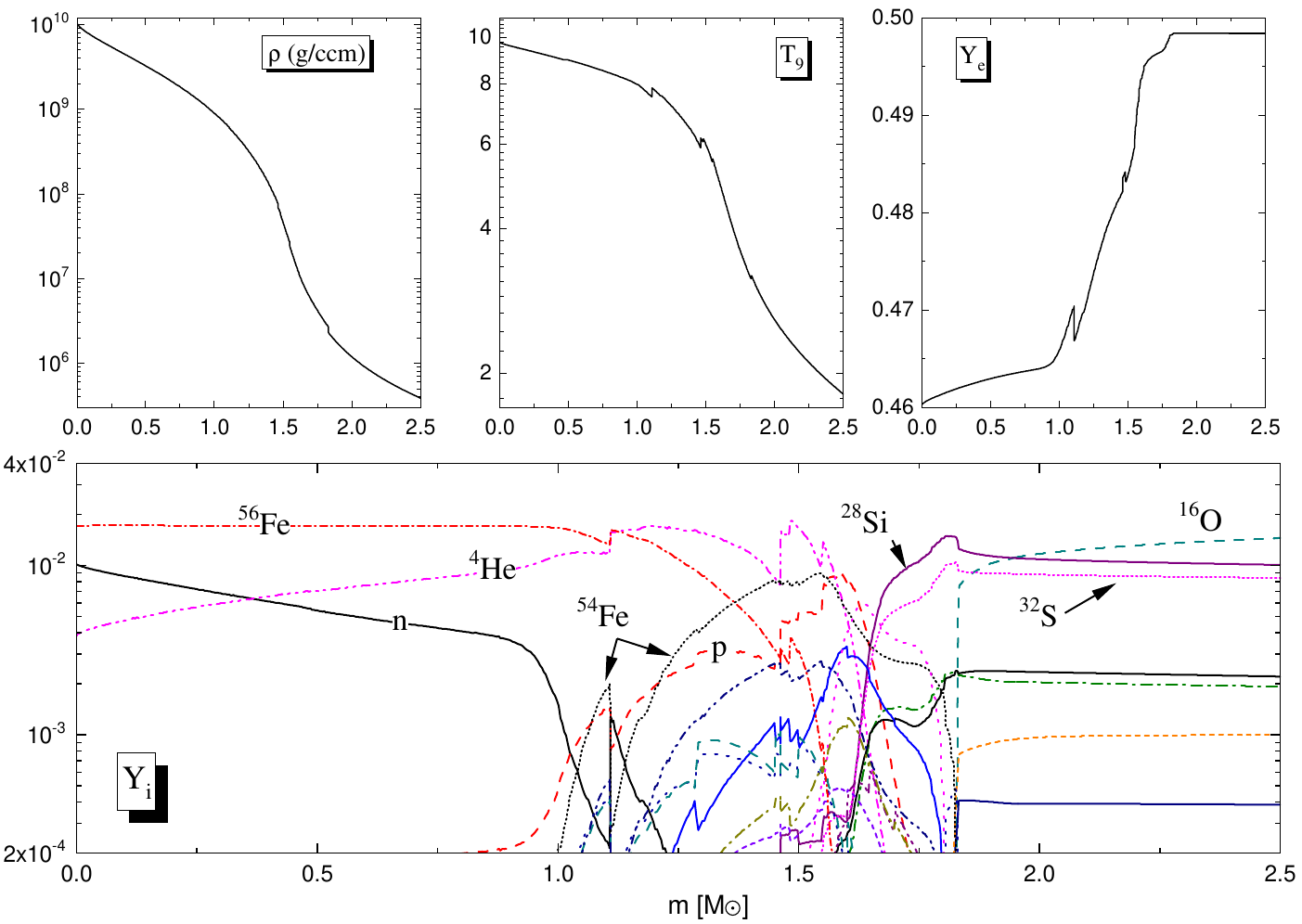}
	\end{center}
\vspace{-0.5cm}
	\caption{Top panels: density $\rho$, temperature $T_9$ and electron fraction $\Ye$ profiles along the mass coordinate $m$  for the \texttt{25\_79\_0p005\_ml} pre-supernova model at the onset of the core collapse. Bottom panel: the chemical composition at the same conditions.} \label{Fig.track_25_79}
\end{figure*}

Let us now consider a characteristic structure of the pre-supernova model we use in the present work. As in our previous study~\cite{Dzhioev_Particles6},  we use the model
\texttt{25\_79\_0p005\_ml} from~\cite{Farmer2016}, which is a realistic pre-supernova model with a good mass resolution and the central temperature which is high enough for our estimates. Its name means that the initial mass of the model was $25 M_\odot$, the nuclear reaction network was \texttt{mesa-79.net}, the maximum mass of a computational cell was $0.005 M_\odot$, and the mass loss during the stellar evolution was taken into account (for more details see \cite{Farmer2016}).
The authors of \cite{Farmer2016} employ the stellar evolution code MESA \citep{Paxton2015}, version 7624. In the output, MESA produces  time-evolving profiles of mass density, temperature, electron fraction $Y_\mathrm{e}$ and chemical composition as  a function of the mass coordinate. The profile that we use corresponds to the onset of the core collapse, which is defined as the time when the infall velocity exceeds 1000 km/s anywhere in the star.

The detailed  structure of this profile is given in Fig.~\ref{Fig.track_25_79}.  Three top  panels show  the density $\rho$, temperature $T_9\equiv T/10^9~K$, and the electron fraction $\Ye$ as a function of the mass coordinate $m$ (in solar masses $M_\odot$) in the central part of the core at $0\leq m\leq 2.5$. The bottom panel shows the dimensionless concentrations of matter components $Y_i=n_i/\WI{n}{b}$, where $\WI{n}{b}$ is the concentration of baryons; $Y_i$ are related to commonly used  weight fractions $X_i$ as $X_i=A_i Y_i$, where $A_i$ is the mass number of the component. As can be seen, the \isn{Fe}{56} isotope is predominate in the hottest part of the core  up to $m\approx1\,M_\odot$.

\section{Neutrino generation inside pre-SN} \label{sec_generation}

In this section, we consider various processes of neutrino generation under the pre-supernova conditions described above. First, we describe the generation processes themselves, then we calculate the corresponding neutrino luminosities inside the stellar core, and finally, we determine the spectra of the neutrino flux leaving the star.

\subsection{Neutrino processes}

In what follows we will distinguish neutrino processes according to the type of neutrinos they produce. In Table~\ref{tab:neutr_proc}, in three rows we collect i) the processes of $\WI{\nu}{e}$ generation, ii) the same for $\WI{\bar\nu}{e}$ and iii) $\nu\bar\nu$-processes, where  neutrinos of all three flavors could be born. Among the latter are the ND pair emission by  hot  $^{56}$Fe, electron-positron pair annihilation (PA), plasmon decay (PL), photo-neutrino production (PH), and bremsstrahlung (BR).  We  note that neutrinos from the recombination  have a  luminosity more than four orders of magnitude less than that from other processes and we do not consider them.

\begin{table}
    \caption{Different (anti)neutrino generation processes considered in the present study. The $x$ label denotes three neutrino flavors, i.e. $x=\mathrm{e},\,\mu,\,\tau$.} \label{tab:neutr_proc}
    \centering
    \begin{tabular}{c|c}
    \hline
      label   & process  \\
      \hline
      EC & $\isn{Fe}{56}+e^{-}\rightarrow \isn{Mn}{56}+\WI{\nu}{e}$\\
      $\beta^+$ & $\isn{Fe}{56}\rightarrow \isn{Mn}{56}+e^{+}+\WI{\nu}{e}$\\
      \hline
      PC & $\isn{Fe}{56}+e^{+}\rightarrow \isn{Co}{56}+\WI{\bar{\nu}}{e}$\\
      $\beta^-$ & $\isn{Fe}{56}\rightarrow \isn{Co}{56}+e^{-}+\WI{\bar{\nu}}{e}$\\
      \hline
      ND  & $\isn{Fe}{56}^{*}\rightarrow \nu_x+\bar{\nu}_x$\\
      PA & $e^{-}+e^{+}\rightarrow \nu_x+\bar{\nu}_x$\\
      PL & $\mathrm{plasmon}\rightarrow \nu_x+\bar{\nu}_x$\\
      PH & $e^{-}+\gamma\rightarrow e^{-}+\nu_x+\bar{\nu}_x$\\
      BR & $e^{-}+(A,Z)\rightarrow e^{-}+(A,Z)+\nu_x+\bar{\nu}_x$\\
      \hline
    \end{tabular}
 \end{table}

Before proceeding with the  calculations of  luminosities and spectra for (anti)neutrinos of different types, let us find the total neutrino luminosities due to various $\nu\bar\nu$ processes listed in Table~\ref{tab:neutr_proc}. For this purpose we use the collection of subroutines by F.\,X.~Timmes\footnote{\url{https://cococubed.com/code_pages/nuloss.shtml}}, who implemented the approximations from~\cite{Itoh1996}. To compute the neutrino luminosity due to the ND process, we  apply the TQRPA method.

\begin{figure}
    \centering
\includegraphics[width=1\linewidth]{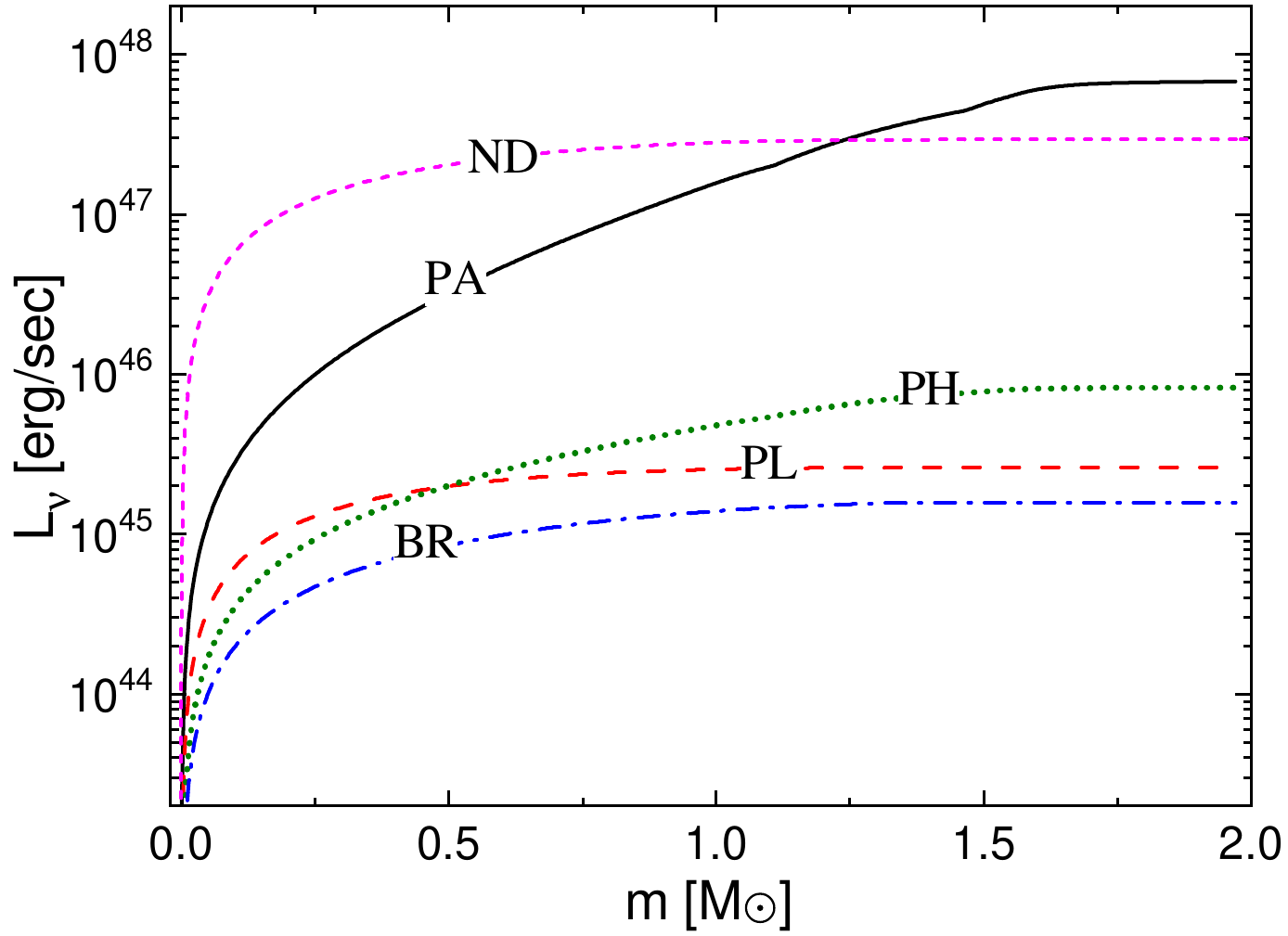}
 \vspace{-0.5cm}
    \caption{Total (anti)neutrino  luminosities inside the central part of the star for the pair production processes listed in Table~\ref{tab:neutr_proc}.}
    \label{fig:Thermo}
\end{figure}

  Figure~\ref{fig:Thermo}  shows the \emph{total}
  (i.e. $\nu$ and $\bar{\nu}$ of all flavors) luminosities in the central part of the star  as a function of the mass coordinate $m$. As seen from the figure, only PA and ND neutrinos are important under the considered conditions.   Therefore, in the subsequent discussion we  account  for only  these  pair production processes. It should be noted that ND neutrinos have the highest luminosity in the very central part of the star,  but the growth of their luminosity is almost completely halted after $m\approx 0.5\,\text{M}_\odot$. The reason for this is that the very central part of the star has a sufficiently high temperature \ad{$T_9\approx 9.0-10.0$} to enable the emission of high-energy $\nu\bar\nu$-pairs via the ND process. These high-energy pairs are produced by the de-excitation of the thermally excited GT$_0$ resonance (see Fig.~4 in \cite{Dzhioev_Particles6} and its discussion). In
  \isn{Fe}{56}, the GT$_0$ resonance is located at about $10$\,MeV and its thermal population rapidly decreases as we move into the colder parts of the star. Because of this, the emission of high-energy (anti)neutrinos  due to the ND  process stops, which leads to stagnation in the luminosity.
  In contrast, the luminosity of PA neutrinos continues to grow up to $m\approx 1.7\,\text{M}_\odot$ and ends up being more intense (see \cite{Misiaszek2006PhRvD}, where the problem with the PA luminosity and spectrum calculations is discussed).

  \subsection{$\WI{\nu}{e}$-type neutrino}

  Now we are ready to calculate neutrino luminosities and spectra for specified processes. First, let us consider the processes that produce electron-type neutrinos, i.e. EC, $\beta^+$, ND and PA (see Table~\ref{tab:neutr_proc}).

\begin{figure}
    \centering
\includegraphics[width=1\linewidth]{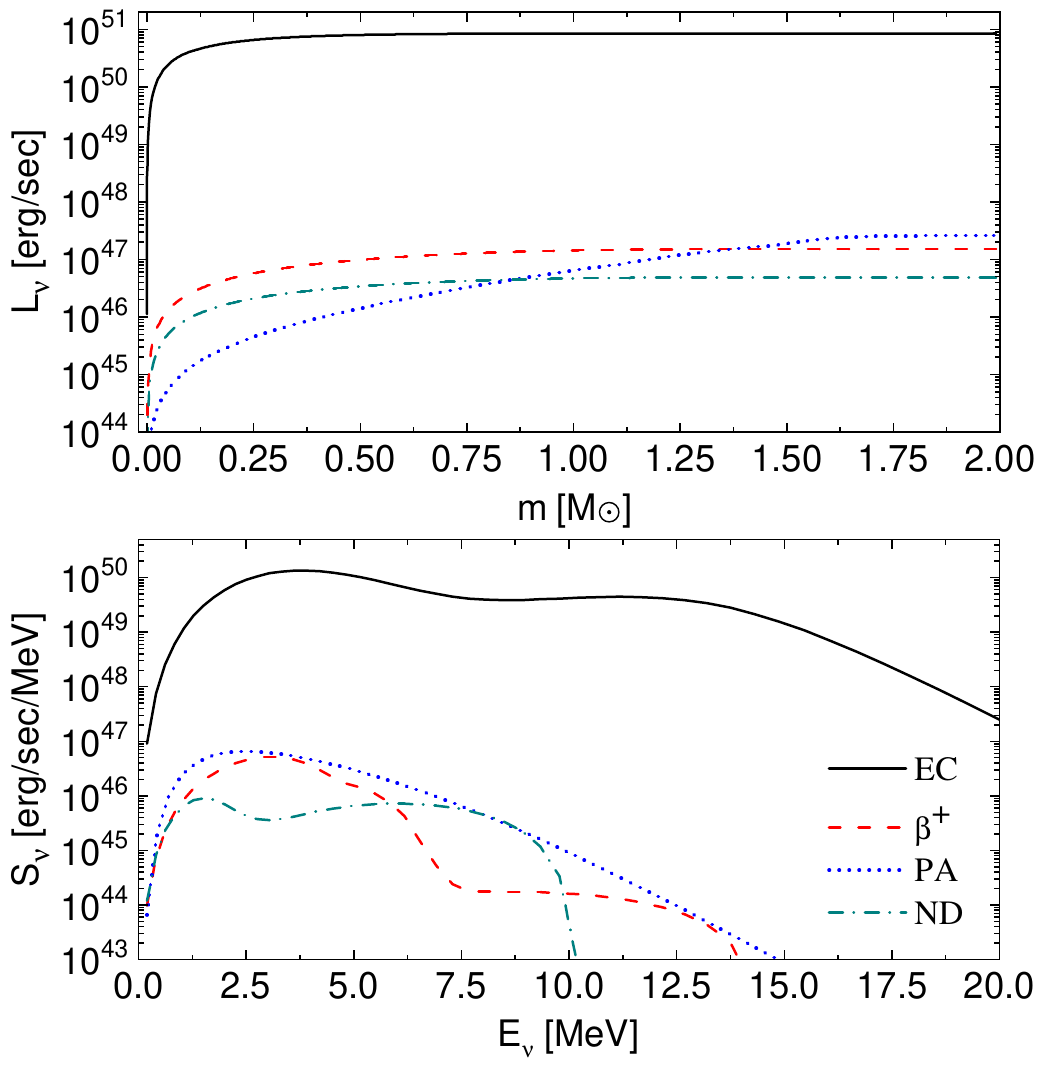}
\vspace{-0.5cm}
    \caption{Upper panel: $\WI{\nu}{e}$-type neutrino luminosities as  functions of \ad{ the mass coordinate} $m$. Lower panel:  spectrum of $\WI{\nu}{e}$-type neutrinos, leaving the star.}
    \label{fig:Enu}
\end{figure}

In Fig.~\ref{fig:Enu}, the $\WI{\nu}{e}$ neutrino  luminosity $L_\nu [\mathrm{erg}/\mathrm{sec}]$ is shown as a function of the mass coordinate in the inner part of a star, $0 \leq m\leq 2$ (upper panel). As is evident from the figure, EC neutrinos are  \ad{several} orders of magnitude more luminous than neutrinos from other processes. \ad{Moreover, the luminosity of EC  neutrinos greatly exceeds the luminosity of $\bar\nu_e$ and  ($\mu,\,\tau$) (anti)neutrinos produced in various nuclear and thermal processes.} Thus, it appears that EC on \isn{Fe}{56} is the dominant energy loss process for the considered pre-supernovae conditions, \ad{and the present study confirms the conclusion made by~\cite{Fischer_PRC88} about the role of electron capture in core-collapse supernovae. }

The lower panel of Fig.~\ref{fig:Enu} shows the \ad{energy luminosity} spectrum $S_\nu [\mathrm{erg}/\mathrm{sec}/\mathrm{MeV}]$ (which is connected to the total  luminosity as $L_\nu=\int S_\nu dE_\nu$) of $\WI{\nu}{e}$ neutrinos leaving the star. It is evident that EC neutrinos not only have a much higher total luminosity \ad{than $\nu_e$ from other processes} but  also a higher average energy (see  the discussion in Section \ref{sec_features}).

\subsection{$\WI{\bar{\nu}}{e}$-type neutrino}

Regardless of the fact that $\WI{\nu}{e}$-type neutrinos, primarily from EC, are the most important energy loss source   in a star, $\WI{\bar{\nu}}{e}$ neutrinos are also important.
This is due to the fact that most detectors are especially sensitive to the inverse beta-decay reaction initiated namely by $\WI{\bar{\nu}}{e}$ (see the discussion in Section \ref{sec_Detection}).

\begin{figure}
    \centering
\includegraphics[width=1\linewidth]{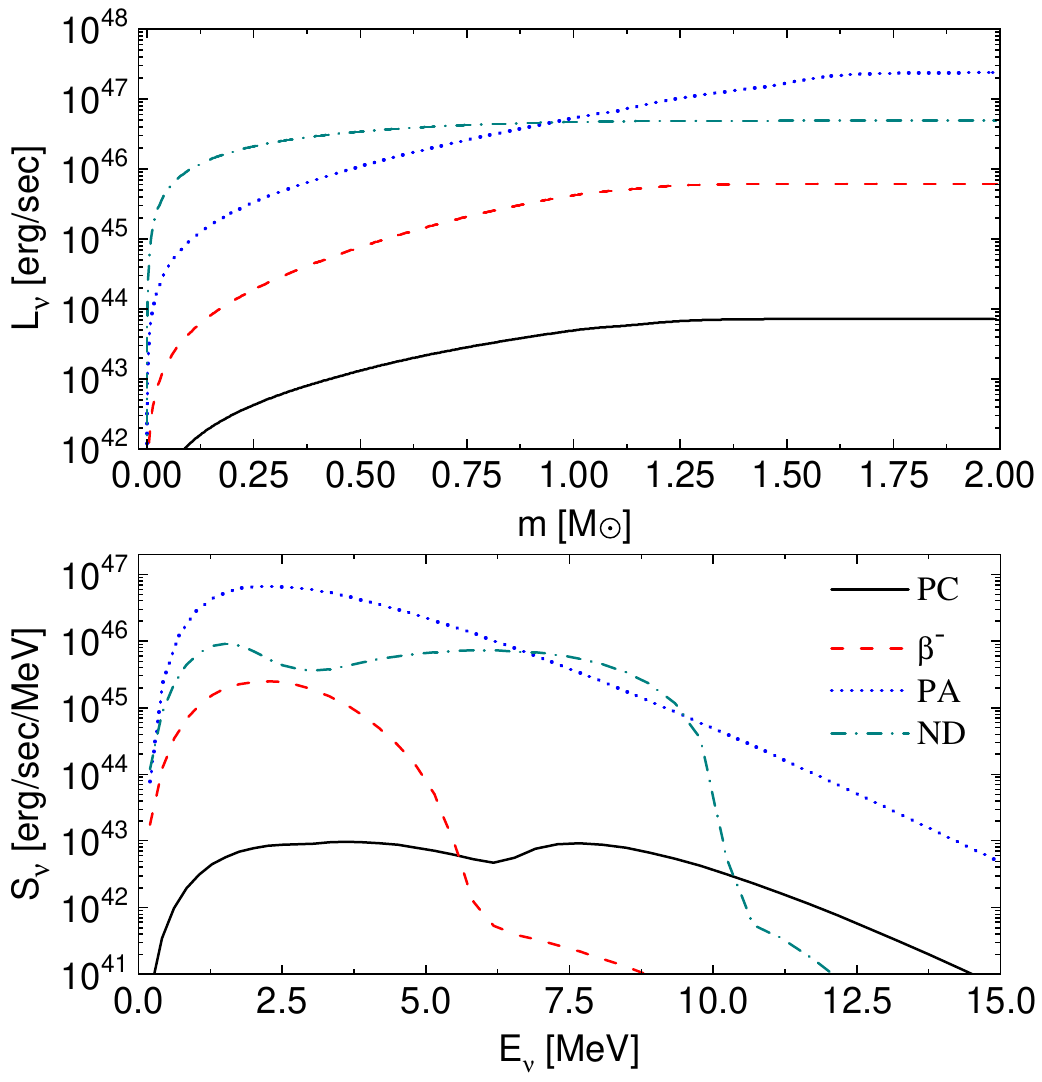}
\vspace{-0.5cm}
    \caption{The same as in Fig.~\ref{fig:Enu} but for $\WI{\bar\nu}{e}$-type neutrinos.}
    \label{fig:AEnu}
\end{figure}

Figure~\ref{fig:AEnu} shows the same quantities (luminosity and spectrum) as in Fig.~\ref{fig:Enu} but for $\WI{\bar{\nu}}{e}$. As obvious from the figure, the essential difference compared to the previously considered case is that there is no one dominant process that produces the $\WI{\bar{\nu}}{e}$ flux. \ad{The PC and $\beta^-$ contributions are small and,} in the very center of the star, the  ND antineutrinos are most luminous. But  as discussed above, due to the temperature decrease as we move from the center and a strong temperature dependence of the ND process, the growth of their luminosity  almost stops completely  after $m\approx 0.5\,\text{M}_\odot$. On the other hand, $L_\nu$ from the PA process grows up till $m\approx 1.7\,\text{M}_\odot$ and the resultant luminosity is several times higher than that  from the ND process. \ad{However, while the spectrum of PA antineutrinos is peaked at low energies,  $\WI{\bar{\nu}}{e}$ from the ND process have a  higher energy  peak in the spectrum at $E_\nu\approx 6.0$\,MeV, and their  luminosity  even exceeds the PA luminosity in the energy range $E_\nu\approx 6-9$~MeV.} We will discuss the implications of this  in Section~\ref{sec_Detection}.

\subsection{$\nu_{\mu,\tau}$- and $\bar{\nu}_{\mu,\tau}$-type neutrino}

\begin{figure}
    \centering
\includegraphics[width=1\linewidth]{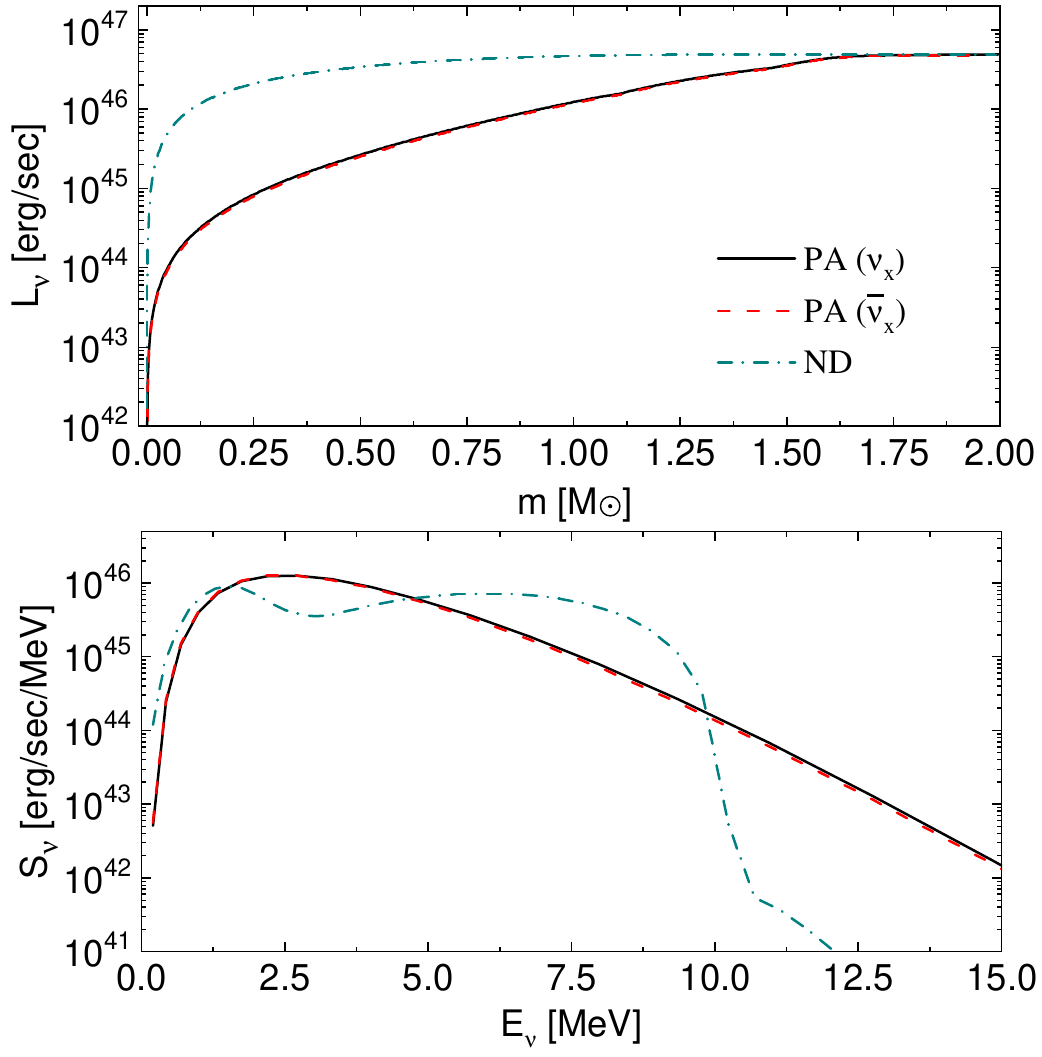}
\vspace{-0.5cm}
    \caption{The same as in Fig.~\ref{fig:Enu} but for   \ad{$\nu_{x}$- and $\bar{\nu}_{x}$}-type neutrinos, \ad{where $x$ stands for $\mu$ or $\tau$.} }
    \label{fig:mut_nu}
\end{figure}

\ad{Despite the fact that emission of heavy-lepton type  (anti)neutrinos   has basically no impact on the energy loss  during the pre-supernova phase, this process might be } important from the point of view of the detection, which is possible due to flavor oscillations.  So we have calculated the luminosities and spectra for \ad{($\mu,\,\tau$)}  (anti)neutrinos as well and put them in Fig.~\ref{fig:mut_nu}.

Referring to the figure, the luminosities and spectra of  $\mu,\, \tau$-type neutrinos and antineutrinos   produced by the pair annihilation are  very close, while for the ND process they are exactly  the same. \ad{Therefore, further we will not distinguish $L_\nu$, $S_\nu$ between $\nu_{\mu,\,\tau}$ and  $\bar\nu_{\mu,\,\tau}$.}  We also see that the final (i.e. at $m=2.0\,\text{M}_\odot$)  PA and ND luminosities are almost the same, but the spectra are quite different. \ad{As in the case of $\bar\nu_e$, the PA spectrum of ($\mu,\,\tau$) (anti)neutrinos is peaked at low energies, while  the  ND spectrum has a high-energy  peak at $E_\nu\approx 6.0$~MeV.  Recall that nuclear de-excitation produces the same spectra for all (anti)neutrino flavors.   When Fig.~\ref{fig:mut_nu} is compared with Fig.~\ref{fig:AEnu}, it is apparent that  production of heavy-lepton flavor (anti)neutrinos in the PA process is lower than that of $e$-type (anti)neutrinos.  This is because  pair annihilation into $\nu_e\bar\nu_e$  proceeds via both neutral and charged currents, while $\nu_{x}\bar\nu_{x}$ ($x=\mu,\,\tau$) pairs are produced only via neutral current $Z^0$ boson exchange~\citep{Misiaszek2006PhRvD}. For this reason, the  ND process appears to be more important for  $\mu$ and $\tau$ (anti)neutrino production  than for $\nu_e$ and $\bar\nu_e$. }

\section{Characteristic features of the TQRPA method}
\label{sec_features}

Before we proceed to discussing the issue of detecting pre-supernova neutrinos on Earth, let us  compare our method with that used in the MESA code.  The MESA code is based on the pre-calculated weak-interaction reaction tables (for EC, PC and $\beta^\pm$ processes), which for \isn{Fe}{56} are obtained applying large-scale shell-model calculations~\citep{Langanke2000NuPhA}. In the present study, the electron (anti)neutrino spectra and luminosities from the LSSM weak-interaction rates are computed on the basis of  extended tables~\footnote{\url{https://theory.gsi.de/~gmartine/rates/}} by applying
 \ad{the  effective $q$-value method~\citep{Langanke_PRC64,Patton2017ApJ_1}. Namely, it is assumed that a weak process on a state in the parent nucleus leads to a single state in the daughter nucleus and the energy difference, $q$, between the initial and final states is the same for all excited states in the parent nucleus (Brink's hypothesis). Then, for a single nucleus  the neutrino spectra from  charge-exchange weak reactions  take the form
\begin{align}\label{effectiveQ}
  &\phi^\text{EC,PC}(E_\nu) = N_\text{EC,PC}\frac{E^2_\nu(E_\nu-q)^2}{1+\exp\left(\dfrac{E_\nu-q\mp\mu_e}{kT}\right)}\Theta(E_\nu-q-m_ec^2),
  \notag\\
  &\phi^{\beta^\pm}(E_\nu) = N_{\beta^\pm}\frac{E^2_\nu(E_\nu-q)^2}{1+\exp\left(\dfrac{E_\nu-q\mp\mu_e}{kT}\right)}\Theta(q-E_\nu-m_ec^2),
\end{align}
where the upper (lower) sign refers to EC and $\beta^+$ (PC and $\beta^-$) and $\mu_e$ is the electron chemical potential.
The effective $q$-value and normalization factors $N_i$ are fitting parameters and they are adjusted to the average (anti)neutrino energy and weak reaction rates
\begin{align}
  &\langle E_{\nu,\bar\nu}\rangle  = \frac{\int^\infty_0(\phi^\text{EC,PC}+\phi^{\beta^\pm})E_\nu dE_\nu }{\int^\infty_0(\phi^\text{EC,PC}+\phi^{\beta^\pm})dE_\nu},
  \\
  &\lambda^i = \int^\infty_0 \phi^i(E_\nu) dE_\nu~~~i=\text{EC,~PC,}~\beta^\pm.
\end{align}
The values of $\langle E_{\nu,\bar\nu}\rangle$ and $\lambda^i$ are listed in the LSSM rate tabulations for a grid of temperature/density points and can be easily interpolated in between \citep{Fuller1985ApJ}.   Note that for  $^{56}$Fe under the considered pre-supernova conditions, the effective $q$-value method results in a negative $q$ for the $\nu_e$ spectrum. This means, that according to Eq.~\eqref{effectiveQ}, only electron captures contribute to  $\nu_e$ production in the charge-exchange channel. In contrast, for the $\bar\nu_e$ spectrum we have $q>m_ec^2$, i.e. both PC and $\beta^-$-decay contribute to $\bar\nu_e$ emission. }

\ad{To compute ND (anti)neutrino spectra within the shell-model method, we use the ground-state GT$_0$ strength distribution $S_{\text{GT}_0}(E)$ for  $^{56}$Fe derived from the LSSM calculations (see Fig.~1 in~\cite{Sampaio_PLB529}). Then, we apply the same method  as in~\cite{Fischer_PRC88} to obtain the GT$_0$ strength function for pair emission. Namely, adopting Brink's hypothesis and exploiting the detailed balance relation~\eqref{DB1}, we get
\begin{align}\label{pair_LSSM}
    \phi^\text{ND}(E_\nu)=\frac{G^2_\text{F} g_A^2}{2\pi^3\hbar^7c^6}\, E^2_\nu
    \int\limits^{\infty}_{E_\nu} S_{\text{GT}_0}(E)\exp\left(-\frac{E}{kT}\right)(E-E_\nu)^2dE.
\end{align}
Combining this expression with the effective $q$-value method for charge-changing processes, we obtain the LSSM (anti)neutrino spectra from nuclear processes. }

\begin{figure}
    \centering
\includegraphics[width=1\linewidth]{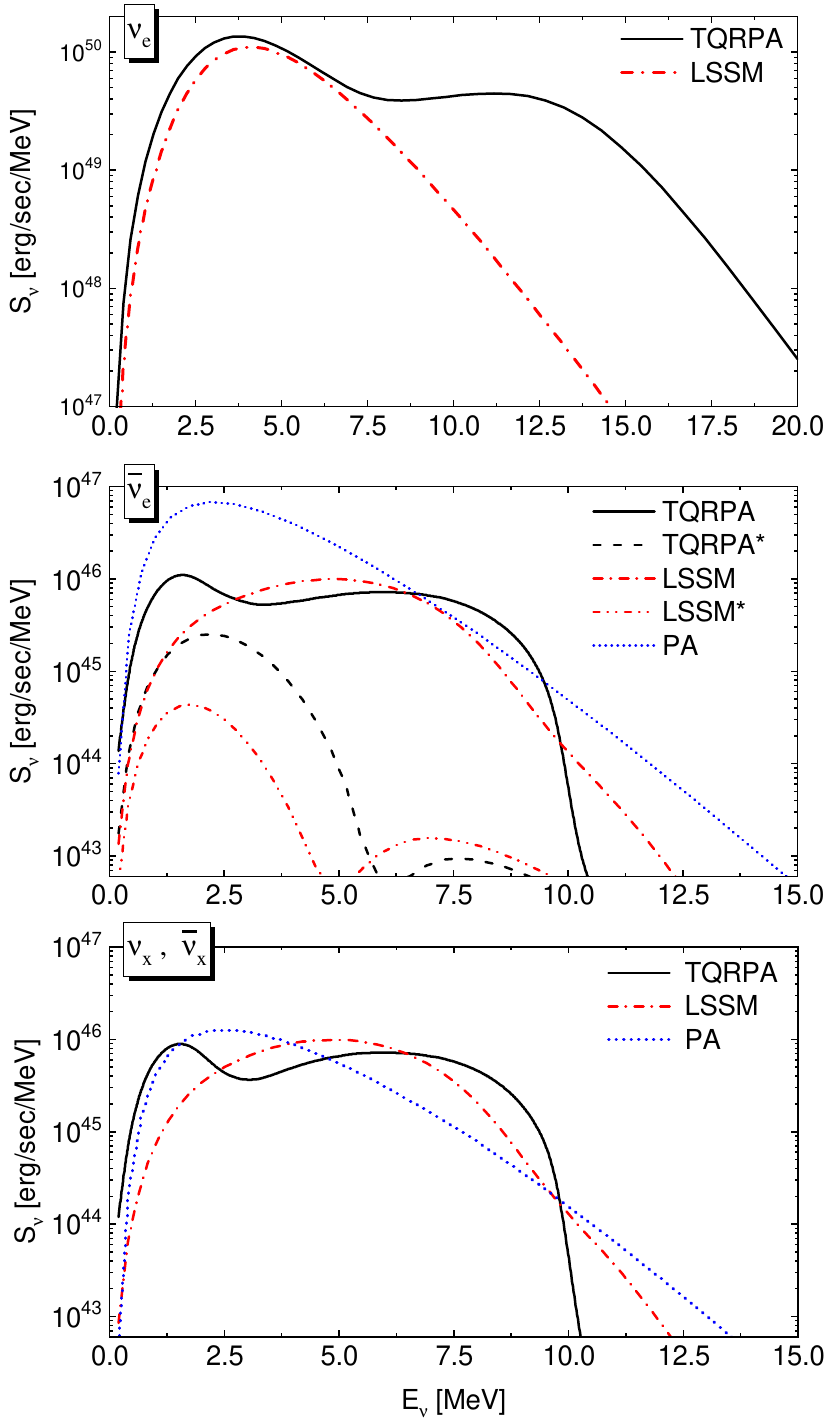}
\vspace{-0.5cm}
    \caption{\ad{Comparison of the TQRPA and LSSM (anti)neutrino spectra from nuclear processes. In both methods, the $\nu_e$ spectra shown in the upper panel are determined almost entirely by the electron capture (see Fig.~\ref{fig:Enu}). In the middle panel, labels "TQRPA"\, and "LSSM"\, represent the $\bar\nu_e$ spectra that include contributions from PC, $\beta^-$-decay and ND, while spectra obtained without the ND contribution are marked as "TQRPA*"\, and "LSSM*".  The $\nu_{x}$ and  $\bar\nu_{x}$ ($x=\mu,\,\tau$) spectra from the ND process are the same in the lower panel.  In the middle and lower panels we also show the spectra from the pair annihilation thermal process.} }      \label{fig:TQRPAvsLSSM_spectra_nucl}
\end{figure}

\ad{In Fig.~\ref{fig:TQRPAvsLSSM_spectra_nucl}, we compare  the TQRPA and LSSM  (anti)neutrino energy luminosity spectra from nuclear processes integrated over the emission region \ad{$m\le 2\, \text{M}_\odot$}.   Focusing on the  upper panel of the figure, we notice  that  both the TQRPA and LSSM calculations give  a low-energy peak in the $\nu_e$ spectrum at $E_\nu\approx4.0$\,MeV. In both methods, EC dominates  $\nu_e$ production and the observed low-energy peak is mainly caused by the GT$_+$ resonance excitation.  The TQRPA approach also produces a second higher energy peak, which makes the spectrum harder. This high-energy peak correspond to $\nu_e$ emitted after  downward negative-energy transitions from thermally excited states. We would like to emphasise that  the absence of the second peak in the LSSM spectrum is the result of the effective $q$-value approximation. In the LSSM calculations of stellar weak-interaction rates, downward transitions are taken into account using the back-resonance method~\citep{Langanke2000NuPhA}, and they produce  a high-energy peak in the neutrino spectrum (see, e.g., Fig.~2 in~\cite{Langanke_PRC64}). }

\ad{The  role of the ND process in the $\bar\nu_e$ spectrum is depicted in the middle panel of Fig.~\ref{fig:TQRPAvsLSSM_spectra_nucl}, where the TQRPA and LSSM spectra with and without the ND contribution are compared with the PA spectrum.   As illustrated,  the TQRPA and LSSM spectra without the ND contribution have a  double-bump structure, reflecting the contribution of PC and  $\beta^-$-decay.  Both methods predict that in the absence of the  ND process, emission of pre-supernova $\bar\nu_e$  is entirely dominated by pair annihilation. Taking nuclear de-excitation into account increases the role of nuclear processes in $\bar\nu_e$ emission.  In both methods, the high-energy ND contribution  around $E_\nu\approx 7-9$\,MeV is comparable or even exceeds the PA contribution. For heavy-lepton flavor (anti)neutrino emission  the role of the ND process is even more important (see the lower panel of Fig.~\ref{fig:TQRPAvsLSSM_spectra_nucl}).  }

\ad{Comparing the ND spectra from the TQRPA and LSSM calculations, we notice that the LSSM  produces a single-peak spectrum, which is different from the double-peak spectrum produced by the TQRPA. In both methods, the high-energy peak corresponds to (ant)neutrinos emitted via de-excitation of the GT$_0$ resonance. The TQRPA calculations of the GT$_0$ strength distribution in $^{56}$Fe predict a somewhat higher centroid energy of the GT$_0$ resonance as compared to the shell-model strength distribution   (see Fig.~1 in~\cite{Dzhioev_PRC94}). As a result, the high-energy peak in the TQRPA spectrum  is slightly shifted to higher energies.  The appearance of a low-energy peak in the  TQRPA spectrum is related to thermally unblocked low-energy GT$_0$ transitions which, owing to the detailed balance principle,   significantly increase the strength  of inverse negative-energy transitions~\citep{Dzhioev_PRC94}. Due to the application of Brink's hypothesis, no such transitions appear in the GT$_0$ strength distribution  obtained from the ground-state LSSM calculations. For this reason, within the TRQPA,  the average energy of emitted ND (anti)neutrinos ($\langle \mathcal E^{\nu\bar\nu}\rangle_\text{TQRPA}=2.97$\,MeV) is lower than that obtained from the LSSM calculations ($\langle \mathcal E^{\nu\bar\nu}\rangle_\text{LSSM}=4.07$\,MeV). However, the luminosities of ND (anti)neutrinos in both methods are nearly the same ($L^{\nu\bar\nu}_\text{TQRPA}=4.9$, $L^{\nu\bar\nu}_\text{LSSM}=4.6$).
It is also interesting to note that $\langle \mathcal E^{\nu\bar\nu}\rangle_\text{TQRPA}$ is close to the values of $\langle \mathcal E^{\nu\bar\nu}\rangle$ obtained in~\cite{Fischer_PRC88} (see Table II) at $T=0.7\,\text{and}\,1.0$\,MeV.   }

\begin{figure}
    \centering
\includegraphics[width=1\linewidth]{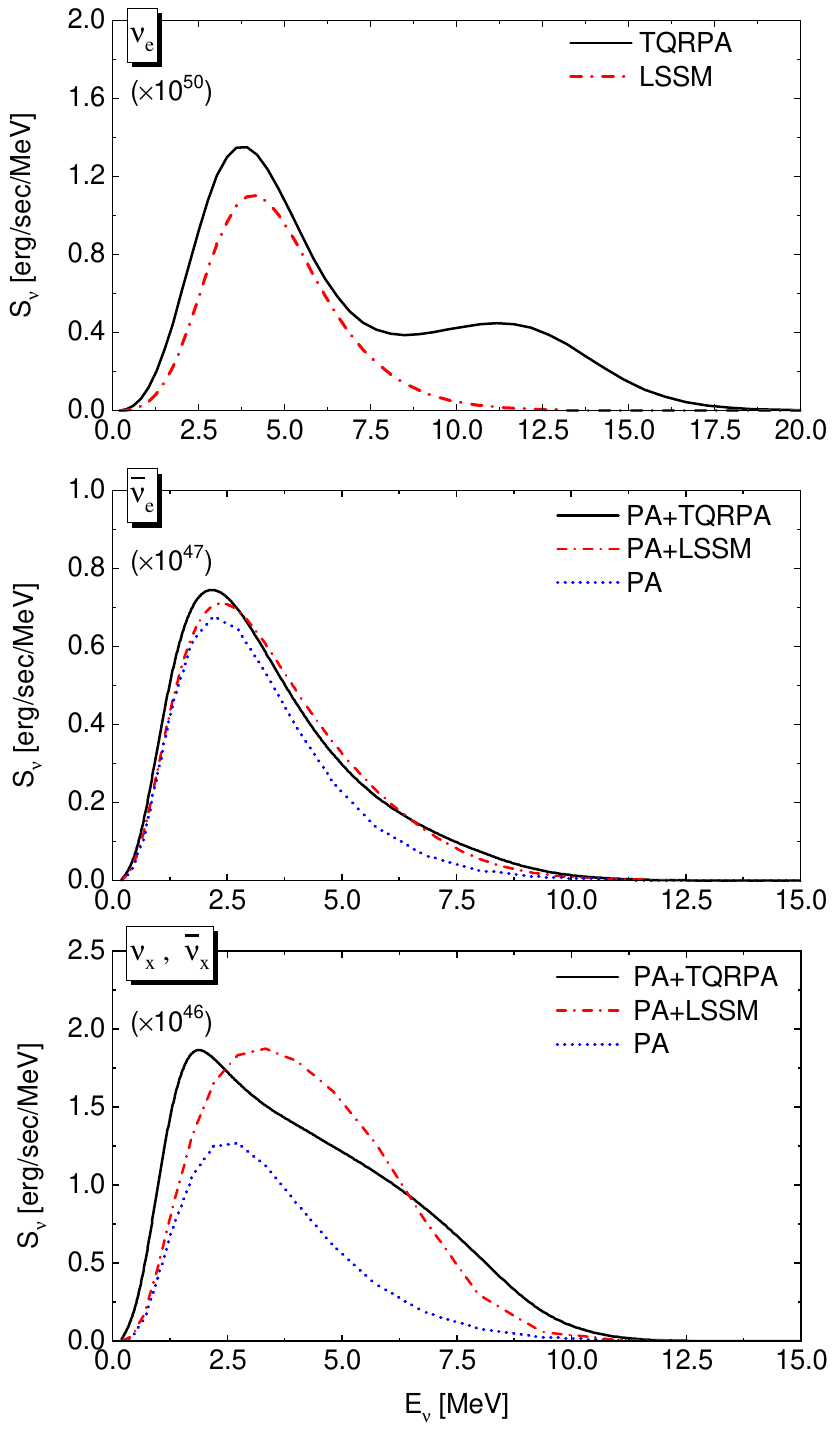}
\vspace{-0.5cm}
    \caption{\ad{Total (i.e. nuclear plus thermal) (anti)neutrino spectra  obtained from the TQRPA and LSSM calculations. For $\nu_e$ neutrinos only electron captures are important and   contributions from other nuclear and thermal processes are negligible.  Note also that the $\bar\nu_e$ spectra without the ND contribution can be considered  with  good accuracy as the PA ones (see the middle panel in Fig.~\ref{fig:TQRPAvsLSSM_spectra_nucl}). The neutrino spectra should be multiplied by the number shown in parenthesis.} }      \label{fig:TQRPAvsLSSM_spectra_total}
\end{figure}

\ad{The  similarities  and differences between the expected total (i.e. nuclear plus thermal) TQRPA and LSSM spectra as well as  the role of the ND process in the pre-supernova (anti)neutrino emission are summarized in Fig.~\ref{fig:TQRPAvsLSSM_spectra_total}.
The comparison  between the TQRPA and LSSM spectra for $\nu_e$  and the discussion of the reasons for their differences has been made above.
Here we just mention that within the TQRPA  the energy luminosity $L_{\nu_e}=\int S_{\nu_e}(E_\nu)dE_{\nu}$ is about twice larger than  that predicted by the LSSM. The reasons for this are discussed below.}

\ad{As for the total $\bar\nu_e$ spectra,  we remind that  in  both methods  the  role of PC and $\beta^-$-decay is negligible and without the ND contribution the spectra are dominated by the PA process. As shown in the middle panel of Fig~\ref{fig:TQRPAvsLSSM_spectra_total}, the ND process  only slightly enhances the emission of low-energy $\bar\nu_e$, but at higher energies $E_\nu=7-9$\,MeV its contribution is comparable or even bigger than the PA one.   It should be emphasised that the increased role of the ND process in the emission of high-energy $\bar\nu_e$ is confirmed by both the TQRPA and LSSM methods.}

\ad{Referring to the lower panel of Fig~\ref{fig:TQRPAvsLSSM_spectra_total}, the dominating role of  ND (anti)neutrinos in the   $\nu_{\mu,\tau}$ and  $\bar\nu_{\mu,\tau}$ spectra is predicted by both the TQRPA and LSSM calculations, and both methods demonstrate the increased contribution of the ND process to the emission of high-energy ($\mu,\,\tau$) (anti)neutrinos.   Thus, the present study generalises the results of~\cite{Fischer_PRC88}
and we can conclude that nuclear de-excitation via $\nu\bar\nu$-pair emission might be a leading source for the production of heavy-lepton type (anti)neutrinos not only during the collapse phase but also during the pre-supernova phase. Once again, we stress  that this result is confirmed by both TQRPA and LSSM calculations. Considering that detection of $\bar\nu_e$ is the main detection channel  and taking into account that flavor  oscillations can swap the $\bar\nu_e$ and $\bar\nu_{\mu,\tau}$ spectra, it is clear that this result might be important for experiments. We will return to this problem in the next two sections.}

\ad{Table~\ref{tab:ME_and_lum} lists the average  (anti)neutrino energies $\langle \mathcal E_{\nu}\rangle=\int S_\nu(E)dE/\int S_\nu(E)E^{-1}dE$ and the luminosities $L_{\nu}=\int S_\nu(E)dE$ computed  from the total spectra shown in Fig.~\ref{fig:TQRPAvsLSSM_spectra_total}. As seen from the table,
both the TQRPA and LSSM calculations predict that at pre-supernova phase the most luminous and  energetic are electron neutrinos. In both methods, the ND process increases the energy luminosities  $L_{\bar\nu_e,\,\nu_{x},\bar\nu_x}$ ($x=\mu,\,\tau$), and this effect is significant for heavy-lepton type (anti)neutrinos.
The ND process also slightly increases the average energy for $\bar\nu_e$, $\nu_x$ and $\bar\nu_x$. This increase is  most pronounced for $\langle \mathcal E_{\nu_x,\bar\nu_x}\rangle_\text{LSSM}$. In the TQRPA, the ND process simultaneously enhances both the low- and high-energy parts of the spectrum (see the lower panel of Fig.~\ref{fig:TQRPAvsLSSM_spectra_total}), which is why $\langle \mathcal E_{\nu_x,\bar\nu_x}\rangle_\text{TQRPA}$ remains   almost unchanged. }

\begin{table}
 \caption{\ad{Average energy $\langle \mathcal E_{\nu}\rangle$ (in MeV) and luminosity $L_\nu$ (in erg/sec) of the (anti)neutrinos produced by all processes listed in Table~\ref{tab:neutr_proc}, for the considered pre-supernova model.  Columns marked  with a star symbol  show data computed  without the ND contribution.}} \label{tab:ME_and_lum}
\begin{tabular}{|c|c|c|c|c}
  \hline\hline
                                      & TQRPA & TQRPA* & LSSM & LSSM* \\
  \hline
  $\langle \mathcal E_{\nu_e}\rangle$          &  4.68  & 4.68    &  4.04 &  4.04  \\
  $L_{\nu_e}$                         &$8.52\times 10^{50}$&$8.52\times
    10^{50}$&$4.67\times 10^{50}$&$4.67\times 10^{50}$   \\
  $\langle\mathcal E_{\bar\nu_e}\rangle$      &   2.49    & 2.41  &  2.58 & 2.41  \\
  $L_{\bar\nu_e}$                     &$2.93\times10^{47}$&$2.44\times10^{47}$&$2.85\times10^{47}$&$2.39\times10^{47}$   \\
  $\langle\mathcal E_{\nu_x,\bar\nu_x}\rangle $& 2.82  & 2.68    &  3.21    & 2.68  \\
  $L_{\nu_{x},\bar\nu_x}$                &$9.78\times10^{46}$&$4.88\times10^{46}$&$9.50\times10^{46}$&$4.88\times10^{46}$   \\
  \hline
\end{tabular}
\end{table}

\begin{figure}
    \centering
\includegraphics[width=1\linewidth]{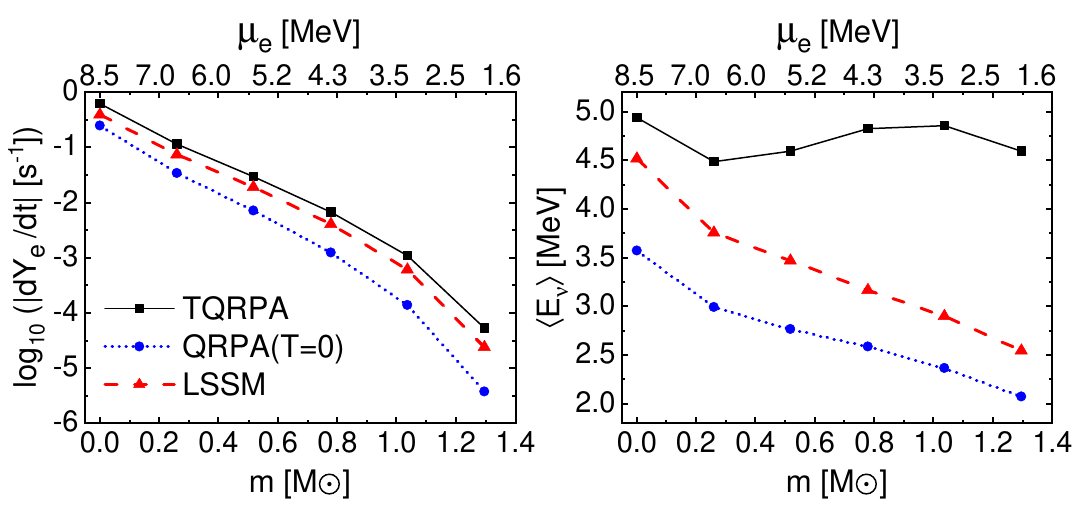}
\vspace{-0.5cm}
    \caption{\ad{Electron capture rate $|dY_e/dt| = Y_{^{56}\text{Fe}}\lambda^\text{EC}$ (left panel) and average energy $\langle E_\nu\rangle$ of emitted EC neutrinos (right panel) as functions of the mass coordinate.  The TQRPA results are compared with the LSSM calculations. To demonstrate the importance of thermal effects, we also show the results of the QRPA calculations, which correspond  to a cold ($T=0$) \isn{Fe}{56} nucleus.  The upper $x$-axis shows the chemical potential $\mu_e$.}}
    \label{fig:TQRPAvsLSSM_ECrates}
\end{figure}

\ad{The TQRPA and LSSM results in Table~\ref{tab:ME_and_lum} are in reasonable agreement with each other except  the electron neutrino luminosity $L_{\nu_e}$. Namely, the TQRPA result is  about twice the luminosity calculated  from the LSSM. We remind that in both methods electron captures on $^{56}$Fe are the main source of $\nu_e$. It must be emphasized that the $L_{\nu_e}$ value obtained in the  LSSM is not a result of the $q$-value approximation, but it can be calculated directly from the LSSM weak-interaction rate tables~\footnote{\url{https://theory.gsi.de/~gmartine/rates/}}.
To understand the reason for the discrepancy, we  note two facts. First,  at pre-supernova temperatures $T_9\approx 2-10$, the TQRPA rates for electron capture by \isn{Fe}{56} generally exceed  the shell-model ones (see Fig.~3  in~\cite{Dzhioev_PRC100_2}). A comparison of the TQRPA and LSSM electron capture rates $|dY_e/dt| = Y_{^{56}\text{Fe}}\lambda^\text{EC}$ along the mass coordinate, shown in the left panel of Fig.~\ref{fig:TQRPAvsLSSM_ECrates}, confirms this observation.
A detailed analysis performed in~\cite{Dzhioev_PRC81} reveals that this disagreement stems from a larger strength of thermally unblocked low- and negative-energy Gamow-Teller transitions predicted by the TQRPA. Second,  a  larger strength of thermally unblocked low- and negative-energy  transitions favours  the emission of  high-energy EC neutrinos. This can be clearly seen (see the right panel of Fig.~\ref{fig:TQRPAvsLSSM_ECrates}) if we compare the TQRPA and LSSM average energy $\langle E_\nu\rangle$ of emitted EC neutrinos. According to the LSSM calculations, $\langle E_{\nu}\rangle$ is significantly lowered when we move from the center of the star  decreasing the chemical potential $\mu_e$, whereas  it varies rather weakly around $\langle E_{\nu}\rangle\approx 4.7$\,MeV within the TQRPA. This stability is a result of the increased fraction of high-energy neutrinos emitted due to thermally unblocked transitions, which compensate for the decrease in available electron energy.   It is clear that both these facts  enhance   the EC neutrino luminosity.}

\ad{To demonstrate the importance of thermal effects, we compute electron capture rates and average neutrino energies within the QRPA, i.e. for a cold nucleus $^{56}$Fe. Referring to Fig.~\ref{fig:TQRPAvsLSSM_ECrates}, without thermal effects, i.e. without thermally unblocked low- and negative-energy GT$_+$ transitions,  EC rates and average neutrino energies  are significantly smaller than  the one  obtained in finite temperature TQRPA and LSSM calculations. We also see that the importance  of thermal effects increases as the chemical potential $\mu_e$ reduces. }

\ad{Obviously,  due to  larger electron capture rates the core should radiate more energy away by neutrino emission, keeping the core on a trajectory with lower temperature and entropy.
Moreover,  larger electron capture rates predicted by the TQRPA for iron-group nuclei should basically result in a faster deleptonization of the progenitor star core at the pre-collapse and  early collapse phases. We note, however, that the difference between the TQRPA and LSSM rates is much smaller than that~\citep{Langanke2000NuPhA} between the LSSM rates and electron capture rates estimated by Fuller, Fowler and Newman on the basis of the independent particle model~\citep{FFN_ApJ25}  . Besides, a larger strength of thermally unblocked negative-energy transitions accelerates the total $\beta^-$-decay rate, which, for a certain range of electron-to-baryon ratios $Y_e$, can exceed the electron capture rate~\citep{GMP_ApJS126} and counteracts the reduction of $Y_e$.  Anyway, definite conclusions about the influence of the TQRPA rates on (pre-)supernova models might be drawn only on the basis (pre)-supernova computer simulations  employing these new rates. }

\section{Oscillations}
\label{sec_oscill}

In principle, the calculated spectra of neutrinos and antineutrinos are not yet sufficient to estimate the effect produced in the Earth's detectors. In fact, we should also take into account the influence of neutrino flavor oscillations. To do this, we will follow a simple approach \citep{Kato2015ApJ}, which allows us to at least qualitatively assess the impact of this effect. We will focus on $\WI{\bar\nu}{e}$-type neutrinos, as being of primary interest in terms of detection (see below Section~\ref{sec_Detection}). So we will assume that the electron  antineutrino spectrum after oscillation can be written as
\begin{equation}
    S_{\bar\nu_e}=p S_{\bar\nu_e}^0+(1{-}p)S^0_{\bar\nu_x}, \label{Osc}
\end{equation}
where index $0$ stands for the original unoscillated   spectra and $x=\mu\,\text{or}\,\tau$. In (\ref{Osc}), the quantity $p$ is the survival probability which is connected to the mixing parameters as $p=\cos^2\theta_{12}\cos^2\theta_{13}\approx 0.676$ for the normal hierarchy (NH) and $p=\sin^2\theta_{13}\approx 0.024$ for the inverted hierarchy (IH) (see, e.g., \cite{Kolupaeva2023}). Thus, the NH provides  moderate mixing while the IH almost suppresses the original $\WI{\bar\nu}{e}$ flux and replaces it by that for $\WI{\bar\nu}{\mu}$ or $\WI{\bar\nu}{\tau}$, both of which have the same spectrum. Let us remind that according to Table~\ref{tab:neutr_proc} and the discussion in Section~\ref{sec_generation}, the original $\WI{\bar\nu}{e}$ flux is  produced by the  PC, $\beta^-$, PA and ND processes, while the  $\WI{\bar\nu}{\mu,\tau}$ flux is formed in the last two  processes.

\begin{figure}
    \centering
\includegraphics[width=1\linewidth]{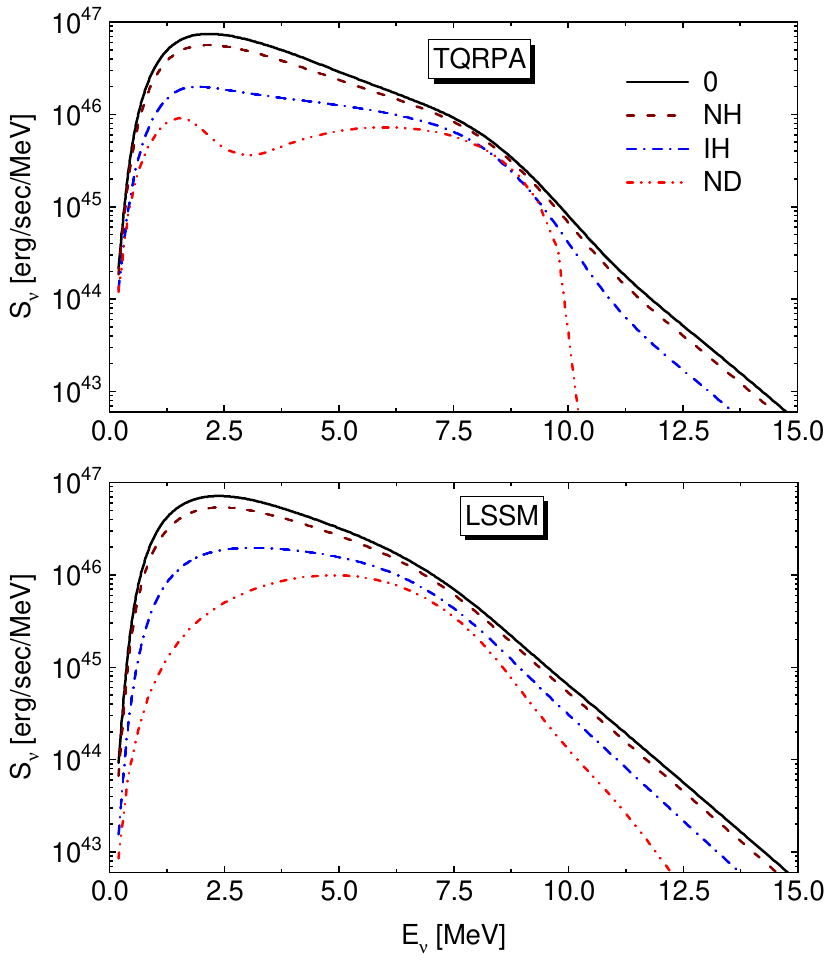}
\vspace{-0.5cm}
    \caption{\ad{The total $\bar\nu_e$ spectra obtained with the normal (NH) and inverted (IH) hierarchy are compared with the original unoscillated spectrum (0). On each plot we also show the ND contribution to the spectra.}}
    \label{fig:Spectra_Oscillation}
\end{figure}

\ad{In  Figure~\ref{fig:Spectra_Oscillation}, we present the total (nuclear plus thermal)  $\bar\nu_e$ spectra for the normal  and inverted   hierarchies  computed within the TQRPA and LSSM approaches and   compare them with  the unoscillated spectra. On each plot we also show the  ND contribution  which does not depend on   oscillations since $S_{\bar\nu_e}^{0,\text{ND}}=S_{\bar\nu_x}^{0,\text{ND}}$.
From the TQRPA and LSSM plots one can see that  NH oscillations  have a small effect on the shape of the original unoscillated spectrum and  just decrease its absolute value. Therefore,  the average energy of $\bar\nu_e$ remains almost unchanged (see Table~\ref{tab:ME_and_lum}) while their total luminosity $L_{\bar\nu_e}$ is suppressed by a factor  of $\sim0.7$. In contrast,  replacing the original $\WI{\bar\nu}{e}$ flux by the  $\WI{\bar\nu}{x}$ one caused by IH oscillations   results in a significant suppression of  the spectrum except the energy range where the ND contribution dominates. As a result, $L_{\bar\nu_e}$ is  lowered by a factor of more than three, while  $\langle\mathcal E_{\bar\nu_e}\rangle_\text{LSSM}$
rises from from 2.6\,MeV to 3.2\,MeV,  and $\langle\mathcal E_{\bar\nu_e}\rangle_\text{TQRPA}$ is practically unchanged.}

\begin{figure}
    \centering
\includegraphics[width=1\linewidth]{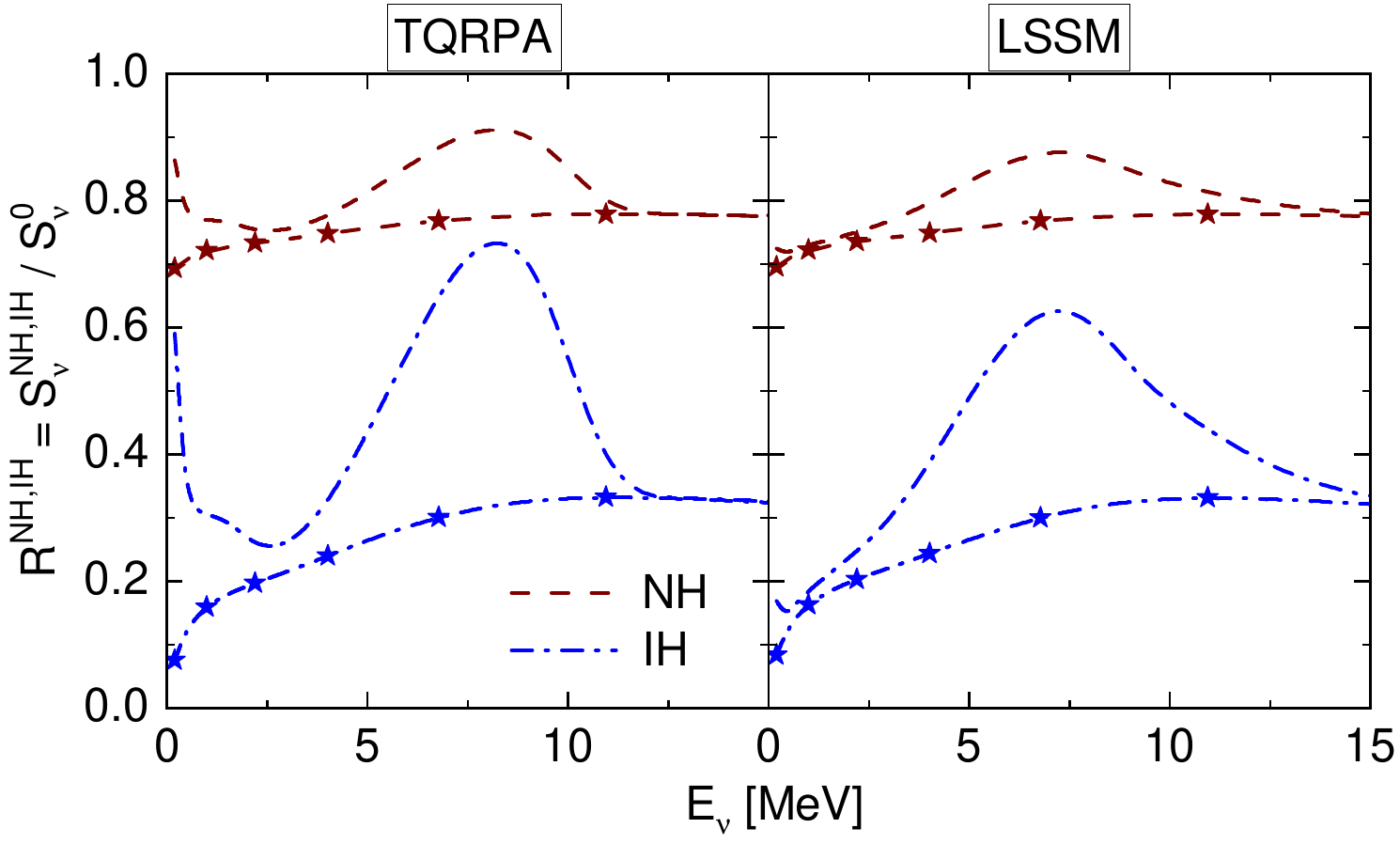}
\vspace{-0.5cm}
    \caption{\ad{The ratio $R^\text{NH,IH}$ of the spectra obtained with the normal and inverted hierarchies to the unoscillated one. The ratios  computed  without the ND contribution are marked with a star symbol. }}
    \label{fig:Ratio_Oscillation}
\end{figure}

\ad{To demonstrate the energy dependence of the $\bar\nu_e$ spectrum suppression due to oscillations, in Fig.~\ref{fig:Ratio_Oscillation} we plot the ratio $R^\text{NH,IH}(E_\nu) = S_{\bar\nu_e}^\text{NH,IH}(E_\nu)/ S_{\bar\nu_e}^\text{0}(E_\nu)$. Referring to the figure, without the ND contribution, the ratio $R^\text{NH}$  is close to the survival probability  $p=0.676$, while the value of $R^\text{IH}$ is determined by the ratio $S^\text{0,PA}_{\bar\nu_x}/ S^\text{0,PA}_{\bar\nu_e}$. The latter increases from 0.1 at low energies up to 0.3 at $E_\nu\approx 15$\,MeV.  The inclusion of the ND process changes a monotonic energy dependence of $R^\text{NH,IH}$. Namely, both the TQRPA and LSSM calculations demonstrate that for high-energy ($E_\nu\approx 5-10$\,MeV) antineutrinos the ND process reduces spectrum suppression  and this effect is most pronounced for  IH oscillations. Within the TRQPA approach, the ND process also increases  the fraction of low-energy neutrinos in the oscillated $\bar\nu_e$ spectrum, which, as  discussed earlier in Section~\ref{sec_features}, is due to temperature-induced increase in  the strength of low-energy GT$_0$ transitions in $^{56}$Fe. }

\section{Detection} \label{sec_Detection}

\begin{figure}
    \centering
\includegraphics[width=1\linewidth]{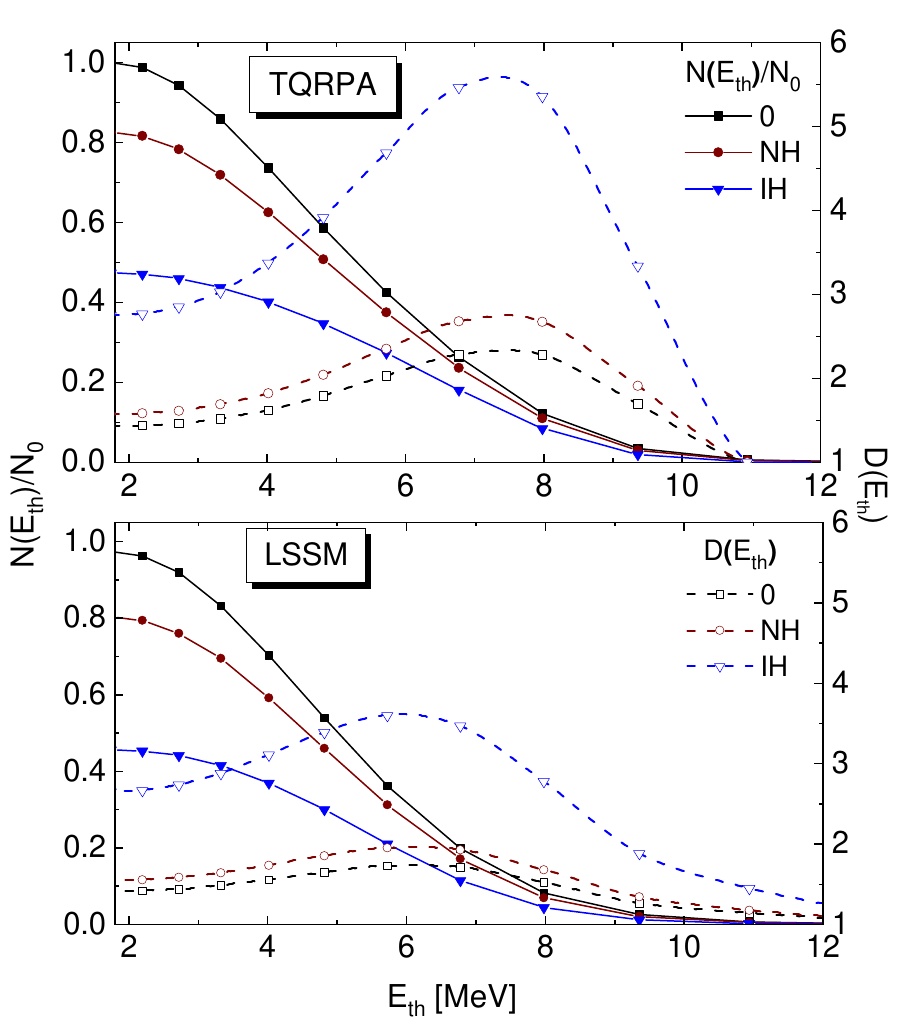}
\vspace{-0.5cm}
    \caption{\ad{The relative number  $N(\WI{E}{th})/N_0$ of events in the detector  and the detection rate  enhancement factor $D(E_\text{th})$  as a function of the threshold energy $\WI{E}{th}$.}  }
    \label{fig:detection}
\end{figure}

The dominant detection process for electron antineutrinos from (pre-)supernova  is the inverse beta-decay:
\begin{equation}
\WI{\bar{\nu}}{e}+p\rightarrow n+e^{+}.
\label{IBD}
\end{equation}
Of course, other reactions  can be important as well (see, e.g., \cite{Manu2022JETP}), but here we will focus on the most standard one.
The cross-section of (\ref{IBD}) can be estimated as (see, e.g., \cite{Odrzhywolek2004APh})
\begin{equation}
\WI{\sigma}{IBD}\sim p_{e^+}E_{e^+},
\end{equation}
where $E_{e^+}=E_{\bar{\nu}}-(\WI{m}{n}{-}\WI{m}{p})c^2$ is the emitted positron energy and $p_{e^+}$ is its momentum. So the minimum threshold  energy required to induce this reaction is $E^{\mathrm{min}}_{\bar{\nu}}=(\WI{m}{n}{-}\WI{m}{p}{+}\WI{m}{e})c^2\approx 1.8$\,MeV. If the detector employed has the detection efficiency 100\% above the  threshold $\WI{E}{th}\ge 1.8$\,MeV, then the number of detected electron antineutrinos with the energy spectrum $S(E_{\bar{\nu}})$ is \ad{expressed as}
\begin{equation}
N(\WI{E}{th})\sim\int\limits_{\WI{E}{th}}^\infty \WI{\sigma}{IBD}(E_{\bar{\nu}})S(E_{\bar{\nu}})\frac{dE_{\bar{\nu}}}{E_{\bar{\nu}}}.\label{N_th}
\end{equation}
Applying this relation and using the \ad{TQRPA and LSSM} spectra shown in Fig.~\ref{fig:Spectra_Oscillation}, we constructed  Fig.~\ref{fig:detection} to compare three cases: NH, IH and unoscillated one.

In Fig.~\ref{fig:detection}, with the \ad{solid} curves \ad{with filled symbols} we denote a relative number of events in the detector, $N(\WI{E}{th})/N_0$,   as a function of the threshold energy. Here the normalisation constant $N_0$ is \ad{$N(\WI{E}{th}=1.8\,\text{MeV})$} for the unoscillated  \ad{TQRPA spectrum.}
\ad{We note that $N^\text{TQRPA}_0$ and $N^\text{LSSM}_0$ are rather close to each other.  As indicated in the figure,  NH oscillations lead to a moderate reduction of the $N(\WI{E}{th})/N_0$ ratio, whereas   IH oscillations reduce
the number of detected events   by about half.}
Besides, the \ad{ratio  $N(\WI{E}{th})/N_0$}  demonstrates  the importance of the detector's threshold parameter $\WI{E}{th}$  and how drastically it can affect  the number of registered pre-supernova $\bar\nu_e$. For example, a rather moderate value of $\WI{E}{th}\approx 5$\,MeV reduces the number of registered events by almost half.

\ad{In order to show the effect of the ND process on neutrino detection, in   Fig.~\ref{fig:detection} we also plot with the dashed curves with empty symbols the quantity
\begin{equation}
D(E_\text{th})=\left(1-\frac{N^\text{ND}(E_\text{th})}{N(E_\text{th})}\right)^{-1},
\end{equation}
which is a detection rate enhancement factor due to  ND neutrinos (right axes). As expected,  ND neutrinos play an important role in the case of   IH oscillations. Namely, for $E_\text{th}\approx 5-8$\,MeV they enhance the TQRPA and LSSM detection rates by factors 5.5 and 3.5, respectively. At low threshold energies the enhancement factor is $D\approx 3.0$. An increase in the survival probability $p$  diminishes the role of ND neutrinos, but even in the case of unoscillated spectra, they  still contribute substantially to $\bar\nu_e$ neutrino detection for $E_\text{th}\approx 5-8$\,MeV. }

Thus, we can conclude that from the viewpoint of detection of pre-supernova electron antineutrinos,  the neutral-current de-excitation of \isn{Fe}{56} is an important process especially in the moments immediately preceding the collapse when the core temperature is high enough for the thermal population of the GT$_0$ resonance.

\section{Summary  and perspectives}

Let us summarize the main results of the presented research. With the TQRPA approach we are able to take into account \ad{thermal effects which influence} various nuclear weak-interaction processes contributing to (anti)neutrino generation under the conditions characteristic for a hot  (pre-)supernova matter.  We focus our attention on  (anti)neutrino production by a hot  \isn{Fe}{56}, which dominates the isotopic composition of the stellar core at  the onset of the collapse.
 The realistic pre-supernova profile we use was obtained with the help of the famous MESA evolutionary code.  In accordance with our previous estimates \citep{Dzhioev_Particles6}, we have found that the TQRPA approach produces not only a  higher energy luminosity of electron neutrinos (mainly born in the EC process), compared to a more conventional \ad{LSSM} approach, but also a  harder neutrino spectrum as compared with those produced with the effective $q$-value approximation. Both these effects are due to a larger strength of  low- and negative-energy GT$_+$ transitions in the thermally excited  \isn{Fe}{56} predicted by the TQRPA.

We also have found that the nuclear neutral-current  de-excitation process via $\nu\bar\nu$-pair emission is at least as important as the electron-positron pair annihilation in the context of the electron antineutrino generation. \ad{Our calculations of $\bar\nu_e$ spectra  based on the  TQRPA and LSSM distributions of the  GT$_0$ strength in $^{56}$Fe have shown that, although $\bar\nu_e$ from the  ND process have a lower total luminosity compared to the PA process,  their luminosity in the high-energy region $E_\nu=7-9$\,MeV is comparable or even exceeds the PA contribution. Moreover, both the TQRPA and LSSM calculations predict that  the ND process is crucial for generation of heavy-lepton flavor (anti)neutrinos under  pre-supernova conditions.} The latter is of extreme importance because of the problem of pre-supernova neutrino registration by the Earth's detectors. We  have also explored the effect of neutrino oscillations and found that ND antineutrinos are important in all cases but especially in the case of the inverted mass hierarchy \ad{when they  increase the detection rate by several times}.

The main goal of the present study is to draw attention to the importance of the careful treatment of neutrino processes with hot nuclei. Our research can be continued and extended in several aspects. First, it is
interesting to study within the TQRPA the effects of nuclear composition on the neutrino luminosities and spectra. For example, according to \cite{Patton2017ApJ_2}, there is no one  dominating nucleus in the center of the star but rather a bunch of iron-group nuclei such as \isn{Ti}{50{-}51}, \isn{Cr}{54} etc. Probably  this  is due to a different and more extended nuclear reaction network than \ad{the one} we use.  We are planning to address this issue in forthcoming studies.

In addition, the diversity of pre-supernovae may be caused by
another factor: a pre-supernova may be part of a close binary system
with mass exchange \citep{Laplace2021A&A} which can lead to significant differences in stellar profiles. And it should be remembered
that not only a core-collapse supernova but also a type
Ia thermonuclear supernova \citep{Odrzywolek2011A&A}  can be a
source of neutrinos that can be detected the on Earth.

\section*{Acknowledgements}

AY and NDB are grateful to S.I.~Blinnikov for useful discussions on neutrino spectra from supernova progenitors on the late stages of their evolution.
AY thanks RSF 22-12-00103 grant for support. \ad{All authors thank the anonymous reviewer for  constructive comments and important suggestions that helped improve this manuscript. }

\section*{Data Availability}

Data generated from computations are reported in the body of the paper. Additional data can be made available upon reasonable request.



\bibliographystyle{mnras}
\bibliography{dzhioev} 





\bsp	
\label{lastpage}
\end{document}